\documentclass[preprint,pre,aps,showpacs]{revtex4}
\usepackage{graphicx}
\begin{document}

\title{Specific heat and bimodality in canonical and grand canonical
versions of the thermodynamic model}
 
\author{G. Chaudhuri \footnote{On leave from Variable Energy Cyclotron Center, 1/AF Bidhan Nagar, Kolkata 700064, India}}
\email{gargi@physics.mcgill.ca}
\affiliation{Physics Department, McGill University, 
Montr{\'e}al, Canada H3A 2T8}

\author{S. Das Gupta}
\email{dasgupta@physics.mcgill.ca}
\affiliation{Physics Department, McGill University, 
Montr{\'e}al, Canada H3A 2T8}

\date{\today}

\begin{abstract}
We address two issues in the thermodynamic model for nuclear disassembly.
Surprisingly large differences in results
for specific heat were seen in predictions from the canonical and 
grand canonical ensembles when the nuclear system passes from
liquid-gas co-existence to the pure gas phase.  We are able to pinpoint
and understand the reasons for such and other discrepancies when they
appear.  There is a subtle but important difference in the physics addressed
in the two models.  In particular if we reformulate the 
parameters in the canonical model
to better approximate the physics addressed in the grand canonical model,
calculations for observables converge.  Next we turn to the issue of
bimodality in the probability distribution of the largest fragment
in both canonical and grand canonical ensembles.  We demonstrate
that this distribution is very closely related to average multiplicities.
The relationship of the bimodal distribution to phase transition is discussed.

\end{abstract}


\maketitle

\section{Introduction}

In models of statistical disassembly of a nuclear system formed by 
the collision of two heavy ions at intermediate energy one assumes
that because of multiple nucleon-nucleon collisions a statistical
equilibrium is reached.  The temperature rises.  The system expands
from normal density and composites are formed on the way to disassembly.
As the system reaches between three to six times the normal volume, the
interactions between composites become unimportant (except for the
long range Coulomb interaction) and one can do a statistical equilibrium
calculation to obtain the yields of composites at a volume called the
freeze-out volume.  The partitioning into available channels can be
solved in the canonical ensemble where the number of particles in the
nuclear system is finite (as it would be in experiments).  
In some experiments, the number of particles can fluctuate around 
a mean value.  In such a case a sum of several canonical calculations
could be appropriate.  Even when the number of particles is fixed
one can hope to replace a canonical model calculation by a 
grand canonical model calculation
where the particle number fluctuates but the
average number can be constrained to a given value.  The case
we will look at corresponds to this situation.  Usually the
grand canonical model is more easily solved.  Hence it is more commonly
used although in the case of nuclear physics (where particle numbers
are typically $\approx$ 200 or less) the use of the canonical ensemble
would be more appropriate. 

Apart from ease of calculation, there is another reason why the grand
canonical model is very useful.  Known properties of nuclear interactions
predict that if nuclear systems were arbitrarily large (consider a fictitious
system where the Coulomb interaction is switched off) disassembly of nuclear
systems would show features of liquid-gas phase transition 
\cite{Dasgupta1}.  Since 
the grand canonical ensemble is expected to become accurate for large
systems, this would seem to be a suitable framework to describe bulk
properties.  The canonical model, built with a constant particle number
in mind, can not be pushed to arbitrarily large particle number although
it can be implemented for fairly big systems containing thousands of
particles.  One can then extraprolate from finite particle number 
systems to the infinite particle number case.  When this was done, a huge
difference showed up between grand canonical and canonical results for
$c_v$ (specific heat per particle at constant volume).  In the grand canonical
model $c_v$ was merely discontinuous at phase transition \cite{Bugaev} but
in the canonical model $c_v$ would go to infinity as the system particle 
number approached infinity \cite{Dasgupta2}.

This discrepancy was examined in detail for a system with 2000 
particles \cite{Das1}.  The analysis showed that in the co-existence region,
even when the average number of particles is 2000, fluctuations in the number
of particles is huge if one uses the grand canonical ensemble.  
The $c_v$ with a fixed number of particles is sharply peaked at a temperature
$T$ where $T$ is a function of the number of particles.  Because in the grand
canonical ensmble the particle number fluctuations are significant when the
average number is 2000, the resulting $c_v$ is smeared out.
What this analysis did not answer are
two significant questions: (a) under what conditions are the grand canonical
results valid and (b) under what conditions can the canonical and 
grand canonical results agree?  We answer both these questions in this work.
The apparent paradox is explained.

We also address the issue of bimodality in the probability distribution of the
largest fragment as a function of the mass number of the largest fragment
for a finite system when the thermodynamic model is
used.  Clearly the canonical model is appropriate here but we also
study bimodality in the grand canonical ensemble, where the average 
value of the number of particles is constrained to a typical value
expected in heavy ion collisions.  In the thermodynamic model it is 
easy to devise systems which may (as in the nuclear matter case) or
may not (i.e., by switching off surface tension
term in binding energy formula) have a phase transition (some other models
may not have this versatility).  Thus in this model the connection of
bimodality to phase transition can be directly established.  It is
interesting to note (as described in detail later) that both canonical model
and the approximation of a finite system with the grand canonical model
show features of bimodality but quantitatively the results are quite
different.

The plan of the paper is as follows.  In section II we set up the
the formulae for the grand canonical ensemble.  In section III the
methodology of the canonical ensemble is presented.  We point out 
that, although not obvious, the two ensembles actually addressed different
physics causing the difference in results for specific heat.  In section IV
we reformulate the canonical model to better approximate the grand 
canonical model.  We show that results for $c_v$
progressively converge.  We then turn to the question of bimodality
in the probability distribution of the largest fragment as a 
function of the fragment mass number.  Formulae
for the probability distribution are given in section V.  In section VI
we present results for this distribution and discuss the bimodality
in both the canonical and grand canonical ensembles.  The connection
between the probability distribution of the largest fragment and
average multiplicity is established in section VII.
Summary is presented in section VIII.

As in \cite{Bugaev,Dasgupta1,Das1} we use one kind of particle and
no Coulomb interaction.  This is adequate for the purpose of this
study and offers considerable numerical simplifications.  Numerous
applications of the canonical \cite{Das2} and the grand canonical models
\cite{Dasgupta3} with two kinds of particles exist where
fits experimental data are the main issues.

\section{Formulae in the grand canonical model}

If we have $n_a$ particles of type $a$, $n_b$ particles of type $b$, $n_c$
particles of type $c$ etc. all enclosed in a volume $V$ and interactions
between particles can be neglected, the grand partition function for
this case can be written as
\begin{eqnarray}
Z_{gr}=\prod_{i=a,b,c...}(1+e^{\beta\mu_i}\omega_i+e^{2\beta\mu_i}
\frac{\omega_i^2}{2!}.......)=\prod_{i=a,b,c,..}\exp(e^{\beta\mu_i}\omega_i)
\end{eqnarray}
Here the $\mu_i$ is the chemical potential and $\omega_i$  the 
canonical partition function of one particle of type $i$.  The average
number of particles of type $i$ is given by $\partial (ln Z_{gr})/
\partial (\beta\mu_i)$ :
\begin{eqnarray}  
n_i=e^{\beta\mu_i}\omega_i
\end{eqnarray}
It is possible that one of the species can be built from two other
species.  In reverse, a heavier species can also break up into two lighter
species.  If $\alpha$ number of particles of type $a$ can combine with
$\beta$ number of particles of type $b$ to produce $\gamma$ 
number of particles
of type $c$, then chemical equilibrium implies \cite{Reif} that the chemical
potentials of $a,b$ and $c$ are related by $\alpha\mu_a+\beta\mu_b=
\gamma\mu_c$.

In our model we have $N$ nucleons in a volume $V$ (which is significantly
larger than the normal nuclear volume) but these nucleons
can be singles or form bound dimers, trimers etc.  Chemical equilibrium implies
that a composite with $k$ bound nucleons has a chemical potential $k\mu$
where $\mu$ is the chemical potential of the monomer (nucleon).  Thus our
ensemble has monomers, dimers, trimers etc. upto some species with $k_{max}$
bound nucleons.  In the actual world of nuclear physics $k_{max}$ terminates
around 250 because of Coulomb interaction but in the model pursued here
we may terminate it arbitrarily at 1 (monomers only), 2(monomers and dimers),
3 or any large $k_{max}$.  It was demonstrated in \cite{Chaudhuri1}
that liquid-gas type phase transition occurs for large $k_{max}>2000$.

The total number of nucleons will be denoted by $N$.  Of course, the
grand canonical ensemble works best when $N$ is very large,
ideally infinite.

We now look into $\omega_i$, the partition function of one composite of $i$
nucleons.This factors into two parts, a traditional translation energy part
and an intrinsic part:
$\omega_i=z_i(tran)z_i(int)$ where
\begin{eqnarray}
z_i(tran)=\frac{V}{h^3}\int \exp(-\beta p^2/2m_i)d^3p=
\frac{V}{h^3}(2\pi m_iT)^{3/2} 
\end{eqnarray}
The intrinsic part $z_i(int)$ of course contains the key to phase transition.
If we regard each composite to exist only in a ground
state with energy 
$e_i^{gr}$, then $z_i(int)=\exp(-\beta e_i^{gr})$.
We use $e_i^{gr}=-iW+\sigma i^{2/3}$ where nuclear physics sets $W$=16 MeV
and $\sigma=18$ MeV.  This simple model itself will
lead to the main results of this paper.  Because of the surface term,
energy per particle drops as $i$ grows.  Let us denote by $F$ the
free energy of the $N$ nucleons where $N$ is the total number of nucleons;
$E$ be the energy and $S$, the entropy: $F=E-TS$.  At finite temperature $F$
will go to its minimum value.  The key issue is how the system of $N$
nucleons breaks up into clusters of different sizes as the temperature
changes.  At low temperature $E$ and hence $F$ minimises
by forming very large clusters (liquid).  But as the temperature increases
$S$ will increase by forming larger number of clusters thus breaking up the
big clusters. Gaseous phase will appear.  How exactly this will  happen
requires calculation and these show that the system goes through a 
first order liquid-gas phase transition \cite{Bugaev,Chaudhuri1}.
As is the common practice, we used here a slightly more
sophisticated model for $z_i(int)$.  We make the surface tension temperature
dependent in conformity with usual parametrisation \cite{Bondorf};
$\sigma(T)=\sigma_0[(T_c^2-T^2)/(T_c^2+T^2)]^{5/4}$.
Here $\sigma_0=$18 MeV and $T_c$=18 MeV.  At $T=T_c$ surface tension 
vanishes and we have a fluid only.  For us this is unimportant as our
focus will be the temperature range 3 to 8 MeV. Also in $z_i$ we include
not only the ground state but also the excited states of the composite
in the Fermi-gas approximation \cite{Bondorf,Dasgupta1}.  The expression
for $z_i(int)$ is now complete and easily tractable. 

Let us now summarise the relevant equations.
For $k=1$ (the nucleon which has no excited states)
\begin{eqnarray}
n_1=\frac{V}{h^3}(2\pi mT)^{3/2}\exp(\mu/T)
\end{eqnarray}
and for $1< k\leq k_{max}$
\begin{eqnarray}
n_k=\frac{V}{h^3}(2\pi mT)^{3/2}k^{3/2}\exp[(\mu k+Wk+kT^2/\epsilon_0-
\sigma(T)k^{2/3})/T]
\end{eqnarray}
Here $n_k$ is the average number of composites with $k$ nucleons.  
In the rest of the paper, for brevity, we will omit the qualifier ``average''.

A useful quantity is the multiplicity defined as
\begin{eqnarray}
M=\sum_{k=1}^{k_{max}}n_k
\end{eqnarray}
The number of nucleons bound in a composite with $k$ nucleons is
$kn_k$ and obviously $N=\sum_{k=1}^{k_{max}}kn_k$.
The pressure is given by
\begin{eqnarray}
p=\sum_{k=1}^{k_{max}}\frac{n_k}{V}T
\end{eqnarray} 
This follows from the identity $pV=TlnZ_{gr}$.

Quantities like $N,V,n_k$ are all extensive variables.  These
equations can all be cast in terms of intensive variables like $N/V=\rho,
n_k/N$ etc so that we can assume both $N$ and $V$ approach very large
values and fluctuations in the number of particles can be ignored.  Thus
for a given temperature and density we solve for $\mu$ using
\begin{eqnarray}
\rho=\frac{(2\pi mT)^{3/2}}{h^3}(\exp(\mu/T)+\sum_{k=2}^{k_{max}}k^{5/2}
\exp[(\mu k+Wk+kT^2/\epsilon_0-\sigma(T)k^{2/3})/T])
\end{eqnarray}
The sum rule $N=\sum_{k=1}^{k_{max}} kn_k$ changes to $1=\sum kn_k/N$.
The energy per particle is given by
\begin{eqnarray}
\frac{E}{N}=\sum_{k=1}^{k_{max}}\frac{n_k}{N}E_k
\end{eqnarray}
where $E_k=\frac{3}{2}T$ for $k$=1 and for $1< k\leq k_{max}$
\begin{eqnarray}
E_k=\frac{3}{2}T+k(-W+\frac{T^2}{\epsilon_0})+\sigma(T)k^{2/3}
-T[\partial\sigma(T)/\partial T]k^{2/3}
\end{eqnarray}
The term $T[\partial\sigma(T)/\partial T]k^{2/3}$ arises from the
temperature dependence of the surface tension.  The effect of this term
is small.  Eqs. (9) and (10) follow from the identity $E=\mu N-
\frac{\partial}{\partial\beta}lnZ_{gr}$.

From what we have described so far it would appear that $V$ in eqs.(3) to
(9) is the freeze-out volume $V$, the volume to which the system
has expanded.  Actually if the freeze-out volume is $V$ then in these 
equations we use $\tilde V$ which is close to $V$ but less.  The reason 
for this is the following.  To a good approximation a composite of
$k$ nucleons is an incompressible sphere with volume $k/\rho_0$
where the value of $\rho_0$ is $\simeq$ 0.16 fm$^{-3}$.   The volume available
for translational motion (eq.(3)) is then $\tilde V=V-V_{excluded}$ where 
we approximate $V_{excluded}\simeq N/\rho_0=V_0$
the normal volume of a nucleus with $N$ nucleons.  Similar corrections
are implicit in Van der Waals equation of state.  This is meant
to take care of hard sphere interactions between different particles.
This answer is approximate.  The correct answer is multiplicity 
dependent.  The approximation of non-interacting composites in a volume
gets to be worse as the volume decreases.  We restrict our
calculation to volumes $V$ greater than $2V_0$.  This is how the calculations
reported proceed.  We choose a value of $V_0/V=
\rho/\rho_0$ from which 
$V_0/\tilde V=\tilde{\rho}/\rho_0=\rho/(\rho_0-\rho)$ is deduced.
This value of $\tilde{\rho}$ is used in eq.(8) to calculate $\mu$
and all other quantities.  We plot results as function of $\rho/\rho_0$.
If we plotted them as function of $\tilde{\rho}/\rho_0$ the plot would
shift to the right.

\section{The canonical model solution}
The statistical equilibrium model as described above can be solved
for a given fixed number of particles 
when the number of particles $N$ is finite.  No spread in the
number of particles, which is inherent in the grand canonical
ensemble, needs to be made.  
Extensive use of the canonical model has been made to fit
experimental data \cite{Das2} so just an outline will be presented
for completeness.  Among other applications, the canonical model
can be used to study finite particle number effects on phase transition
characteristics.

Consider again $N$ identical particles in an enclosure $V$ and temperature
$T$.  These $N$ nucleons will combine into monomers, dimers, trimers etc.
The partition function of the system in the canonical ensemble
can be written as
\begin{eqnarray}
Q_N=\sum\prod_i\frac{(\omega_i)^{n_i}}{n_i!}
\end{eqnarray}
Here $\omega_i$ is the one particle
partition function of a composite which has $i$
nucleons.  We already encountered $\omega_i$ in section II: 
$\omega_i=z_i(tran)z_i(int)$
with $z_i(tran)$ and $z_i(int)$ given in detail.
Other forms for $\omega_i$ can be used in the method outlined
here.  The summation in eq.(11) is over all partitions which satisfy
$N=\sum in_i$.  The summation is non-trivial as the number of partitions which
satisfy the sum is enormous.  We can define a
given allowed partition to be a channel.  The probablity of the occurrence
of a given channel $P(\vec n)\equiv P(n_1,n_2,n_3....)$ is
\begin{eqnarray}
P(\vec n)=\frac{1}{Q_N}\prod\frac{(\omega_i)^{n_i}}{n_i!}.
\end{eqnarray}
The average number of composites of $i$ nucleons is easily seen from
the above equation to be
\begin{eqnarray}
n_i= \omega_i\frac{Q_{N-i}}{Q_N}
\end{eqnarray}
Since $\sum in_i=N$, one readily arrives at a recursion relation
\cite{Chase}
\begin{eqnarray}
Q_N=\frac{1}{N}\sum_{k=1}^{N}k\omega_kQ_{N-k}
\end{eqnarray}
For one kind of particle, $Q_N$ above is easily evaluated on a computer for
$N$ as large as 3000 in matter of seconds.  It is this recursion relation
that makes the computation so easy in the model.  Of course, once one has
the partition function all relevant thermodynamic quantities can be
computed.  For example, eq. (7) still gives the expression for pressure
although one could correct for the center of mass motion by reducing
the multiplicity by 1: $p=T(M-1)/\tilde V$.  The chemical potential
can be calculated from $\mu=F(N)-F(N-1)$ where the free energy is
$F(N)=-T~ln~Q_N$ which is readily available from the calculation.

\section{Generating grand canonical results from the canonical ensemble}
We first consider pressure (eq.(7)) in the grand canonical ensemble.  The
$V$ in eq.(7) cancels out the $V$ in eqs. (4)and (5) and thus pressure
is given in terms of intensive variables directly.  We may assume that
this is truly the pressure in infinite systems ($V$ and $n_k$ arbitrarily
large in which case fluctuations in the grand canonical ensemble can
be ignored).  However the grand canonical answer does depend upon the
value of $k_{max}$.  In Fig.1 we have used $k_{max}$=2000, a value large enough
so that liquid-gas transition type features emerge (the flatness of 
pressure against $\rho$).  For $k_{max}$ significantly lower, the
flatness disappears (see details in \cite{Chaudhuri1}).  In Fig.1 we
also show several canonical model results all with the same $k_{max}$=2000
but different values of $N$.  For $N$=2000, the canonical results are 
quite different  from the grand canonical results
except for very low densities.  In particular a region
of mechanical instability is seen which can give rise to a region of 
negative $c_p$, the specific heat per particle at constant pressure
(see \cite{Das2} for detailed discussion).  In the same figure, we
have also shown prssures in the canonical model when $N$=100,000 
and 500,000.  We see that the pressure approaches the grand canonical
value as $N$ increases (the periodicity obvious in the curve for $N$=
100,000 arises from the fact the largest composite has $k=$2000 but
we will not get into a detailed analysis here).  The conclusion here is that
the grand canonical value of pressure in Fig.1
refers to a system which has
$N=\infty$ where the largest cluster has $k=2000$.  This is quite different
from the usual canonical model result which would have $N=k_{max}=2000$.
To address the physics of the grand canonical model we keep $k_{max}$
still at 2000 but need to keep on increasing the value of $N$.  Then the
canonical results converge towards the grand canonical values.

For this given problem we have approached the
grand canonical result
from a canonical ensemble.  One can consider the reverse problem:
getting the canonical result starting from the grand canonical model.
It is of course obvious that the correspondence would be exact provided
one uses the appropriate $V$ in the grand canonical ensemble 
and then projects from it the part which has an
exact $N$.  This is because the grand canonical ensemble
is a particular weighted sum of canonical ensembles with different $N$'s.

Let us now turn to Fig.2 which deals with $c_v$, the specific heat per
particle at constant volume.  We again keep $k_{max}=2000$.  One finds that 
the $c_v$ in a canonical calculation for $N=k_{max}=2000$ produces a very
sharp peak.  The grand canonical expression for energy per particle
(eq.(9)) is an intensive quanity and so is its derivative $c_v$.
We expect this grand canonical result for $c_v$ is valid for $N$
very large.  Comparison shows that the grand canonical result differs
drastically from the canonical $N$=2000 result in a very narrow
window when the system passes from the co-existence to a pure gas phase.
Can we make the results converge by successively increasing the value of 
$N$ in the canonical model?  The answer is ``yes'' as Fig.2 demonstrates.
We see that the canonical result with $k_{max}$=2000 approaches
the grand canonical result with $k_{max}$=2000 as the number of particles
$N$ in the canonical calculations is progressively increased beyond
$N$=2000.

To summarise: the grand canonical model is applicable when $N>>k_{max}$.
The limit $N>>k_{max}; k_{max}\rightarrow\infty$ is robust (as shown in
\cite{Chaudhuri1}) and produces a first order phase transition.  This model
is distinct from the canonical model $N=k_{max}; k_{max}\rightarrow\infty$.  
There is no scaling in this latter model:$N=k_{max}$, both $N$ and $k_{max}$
very large is not equivalent within a factor of scaling to 
a system with $2N=2k_{max}$.
We are unable to provide a robust limit for the canonical
model of $N=k_{max}; N\rightarrow\infty$.  Fig. 3 shows the progression 
of the $E/N$ and pressure curves as $N=k_{max}$ increases from 2000 to
50,000. 

\section{Bimodality: the basic formulae in grand canonical and canonical
ensemble}
In event by event analysis in experiments,
one can in principle ascertain the largest mass
(or the largest charge) emerging in each event from multifragmentation.  
The probability
distribution of this largest mass can be plotted as a function of the value
of mass of this largest fragment.  It is shown that a bimodality in this 
distribution at a  certain temperature is a signature  
of a first order phase transition: that is, if the system were infinitely 
large it would have a first order phase transition
\cite{Gulminelli,Pichon}.  Thus from a finite
system one can have a signal for phase transition.  We will now see
how the probability distribution of the largest fragment as a function of the
mass of the largest fragment can be computed in the thermodynamic model in the
two ensembles.  First the grand canonical ensemble.

The grand canonical ensemble works best for a large system and we have
already seen in the previous section that application of this model to
multifragmentation of finite nuclei can lead to serious errors in some
temperature (equivalently energy) window.  Nonetheless, let us proceed
to see how results can be derived.  We fix a value for $\rho/\rho_0$
(in figs. (4) and (5) we have kept this at 0.25) and choose the
appropriate value of the volume so that the average number 
$N=\sum_1^{k_{max}}kn_k$
is 150, the system whose results we show.  The heaviest composite allowed 
in the model $k_{max}$ is also 150.

From eqs. (1) and (2) one can derive
that the probability that a particular composite with $k$ nucleons does
not occur at all is

$\frac{1}{e^{n_k}}$ 

and the probability that it occurs
at least once or more is

$\frac{e^{n_k}-1}{e^{n_k}}$.

Note that our $n_k$ here is the same as $n_k$'s of eqs. (4) and (5),
the average values in the grand canonical ensemble. The probability that
$k$ is the highest mass fragment in an event is then given
by ($k< k_{max}$)
\begin{eqnarray}
P_m(k)=\frac{e^ {n_k}-1}{e^{n_k}}e^
{-(n_{k+1}+n_{k+2}+.....n_{kmax})}
\end{eqnarray}
From the above eq. one readily derives
\begin{eqnarray}
\frac{P_m(k+1)}{P_m(k)}=\frac{e^{n_{k+1}}-1}{e^{n_k}-1}e^{n_k}=
\frac{e^{n_{k+1}}-1}{1-e^{-n_k}}
\end{eqnarray}
If $n_{k+1}>n_k$ then $P_m(k+1)>P_m(k)$. 
If further both $n_{k+1}$ and $n_k$ are small compared to 1 then
\begin{eqnarray}
\frac{P_m(k+1)}{P_m(k)}=\frac{n_{k+1}}{n_k}
\end{eqnarray}
Let us turn to the calculation of the probability disribution of
the largest fragment as a function of the mass of the largest fragment
in the canonical model.  A detailed formulation when two kinds of
particles are present was given in a recent paper \cite{Chaudhuri2}
but for competeness, we review the development.There is an enormous number of
channels in Eq.(11).  Different channels will have different values 
for the largest fragment.  For example there is a term $\frac{\omega_1^N}{N!}$
in the sum of Eq.(11).  In this channel all the fragments and hence
also the largest fragment has mass 1.  
The probability
of this channel occurring is (from Eq.(12)) 
$P_m(1)=\frac{1}{Q_N}\frac{\omega_1^N}{N!}$.
The full partition function can be written as
$Q_N=Q_(\omega_1,\omega_2,\omega_3,.......\omega_{kmax})$.
If we construct a $Q_N$
where we set all $\omega$'s except $\omega_1$
to be zero then this 
$Q_N(\omega_1,0,0,0..............)
=\frac{\omega_1^N}{N!}$ and this has the largest mass 1.
Consider now constructing a $Q_N$ with only two $\omega$'s
: $Q_N(\omega_1,\omega_2,0,0,0,......)$.
This will have the largest mass sometimes 1 
(as $\frac{\omega_1^N}{N!}$ is still there)
and sometimes 2 (as, for example, in the term
$\frac{\omega_2^3}{3!}\frac{\omega_1^{N-6}}{(N-6)!})$.  It then follows that
\begin{eqnarray}
P_m(k)=\frac{Q_N(\omega_1,\omega_2,...\omega_k,0,0,0..)-Q_N(\omega_1,\omega_2,
...\omega_{k-1},0,0,0...)}{Q_N}
\end{eqnarray}
In the above the first term in the numerator takes care of the occurrence
of all partitions where the largest fragment is between 1 and $k$ and the
second term takes care of all the partitions where the largest fragment is
between 1 and $k-1$.  The difference, divided by $Q_N$ is the desired answer.

Since one has the general formula for the probability $P_m(k)$, one can
compute the average value of the mass of the largest fragment as well
as the root mean square deviation.  In fact, these have been measured
in some experiments \cite{Elliott} and have recently been calculated
\cite{Chaudhuri2}.  But we will not need this for this paper.

\section{Representative Results}
The probability distribution $P_m(k)$ of the largest fragment as a
function of $k$ where $k$ is the largest fragment in an event is
shown in Fig.(4) where the freeze-out density $\rho/\rho_0$ is 0.25
and the dissociating system has $N$=150 (for the grand canonical ensemble
the average value is 150).  The canonical and grand canonical results 
are quite different but both display bimodality (there are two
maxima with similar heights), the grand canonical at 
temperature $\approx 5.9$ MeV and the canonical at temperature $\approx 6.2$
MeV.  In Fig.(5) we have compared the $n_k$'s of the two models.  Near
the end value 150 the differences are very substantial at all
temparatures.  At lower values of $k$ they agree very well at $T=$
6.8 MeV, quite well at $T$=6.2 MeV but gets worse at lower temperatures
becoming quite different at $T$=5.0 MeV.  These differences have been noted
and discussed before \cite{Das3}.

\section{Connection between probability distribution of the largest
fragment and average multiplicity}
The very first experiments in heavy ion collisions measured $n_k$, 
the average multiplicity against $k$.  One of the earliest postulates
was the following.  At low energy $n_k$ first falls with $k$ but 
after reaching a minimum rises again.  This is the so-called ``U'' shape.
This shape at lower temperature is an indication that the system will 
undergo a liquid-gas type phase transition.
As the energy of collision increases, the height of the maximum on the
heavier side will decrease, will then disappear (this marks the phase
transition temperature).  At higher
energy, $n_k$ decreases monotonically with $k$.
This is discussed in many places including \cite{Dasgupta1,Dasgupta2}.
Basically then one looks at the behaviour of $n_k$ as a function of
$k$ and energy as one signature of phase transition.  Since bimodality
in the probability distribution is also a signature of phase transition,
we hope to get a connection between $P_m(k)$ and $n_k$.

For bimodality one requires that after the minimum following the first maximum,
$P_m(k)$ will rise again witk $k$.  Similarly in conjectures involving
the multiplicity, $n_k$, after reaching a minimum must rise again with $k$.
These two features are intimately related.  In the grand canonical
model this is very simple to prove.  Equation (16) shows that if
$n_{k+1}>n_k$ then $P_m(k+1)>P_m(k)$ and bimodality can happen.  The
reverse is not true; $n_{k+1}<n_k$ does not imply that $P_m(k+1)$ is
less than $P_m(k)$.

There is similar connection in the canonical model.  
Here it can be proven that on the heavier side $k>N/2$, a rise
of $n_k$ with $k$ guarantees that $P_m(k)$ will rise with $k$.  In fact
it is even more direct than that.  For $k>N/2$, we have an
equality; $P_m(k)=n_k$.  This can be proven from eq. (18) but there is a
an easier proof.  We can rewrite $P_m(k)$ as a sum of terms:
\begin{eqnarray}
P_m(k)=P_m^1(k)+P_m^2(k)+P_m^3(k)+........
\end{eqnarray}
where in each of the terms in the right hand side $k$ is the highest
mass that occurs but in $P_m^1(k)$ the composite $k$ occurs only once,
in $P_m^2(k)$ it occurs twice, in $P_m^3(k)$ it occurs three times and
so on.  Specifically,

$P_m^1(k)=\frac{1}{Q_N}\omega_k\prod_{i<k}\frac{(\omega_i)^{n_i}}{n_i!};~~~~~ 
N-k=\sum_{i=1}^{k-1}in_i$ 

$P_m^2(k)=\frac{1}{Q_N}
\frac{(\omega_k)^2}{2!}\prod_{i<k}\frac{(\omega_i)^{n_i}}
{n_i!};~~~~~N-2k=\sum_i^{k-1}in_i$

It is clear what the structures for higher terms in the series will be.
It is then also obvious that
\begin{eqnarray}
n_k=P_m^1(k)+2P_m^2(k)+3P_m^3(k)+.......+O.C.
\end{eqnarray}
In the above, $O.C.$ stands for other channels where mass $k$ occurs
but it is not the highest mass in the channel.
If $k>N/2$ then only $P_m^1(k)$ exists.  Thus for $k>N/2$ we have
$P_m(k)=n_k$.  If $n_k$ rises with $k$ in this region then so does
$P_m(k)$.  For $k\leq N/2$, the relationship is $n_k\geq P_m(k)$,
with $n_k$ usually significantly larger than $P_m(k)$.  
For bimodality to exist
we need to have $n_{k}$ rising with $k$ in some region $k>N/2$.

The relation $P_m(k)=n_k$ for $k>N/2$ is not limited to the thermodynamic
model only.  It is true in any number conserving model.

\section{Summary}
This paper had two goals.  One, to resolve and understand the difference
between grand canonical and canonical values of specific heat 
and pressure in
thermodynamic models as applied to heavy ion collisions.  This issue
we believe is resolved.  Second, to understand the link between
bimodality in the distribution of the heaviest fragment and the
average multiplicity of fragments (which has also been linked with
aspects of phase transition).  We think we have gained an understanding.
In a later publication we expect to show more results for bimodal
distributions for realistic cases with two kinds of particles and
the Coulomb interactions included.  Calculations for a particular
case have already appeared \cite{Chaudhuri2}.
 
\section{Acknowledgement} 
This work is supported by the Natural Sciences and Engineering Research 
Council of Canada.

\pagebreak

\begin{figure}[htb]
\begin{center}
{\includegraphics[angle=0,width=15cm]{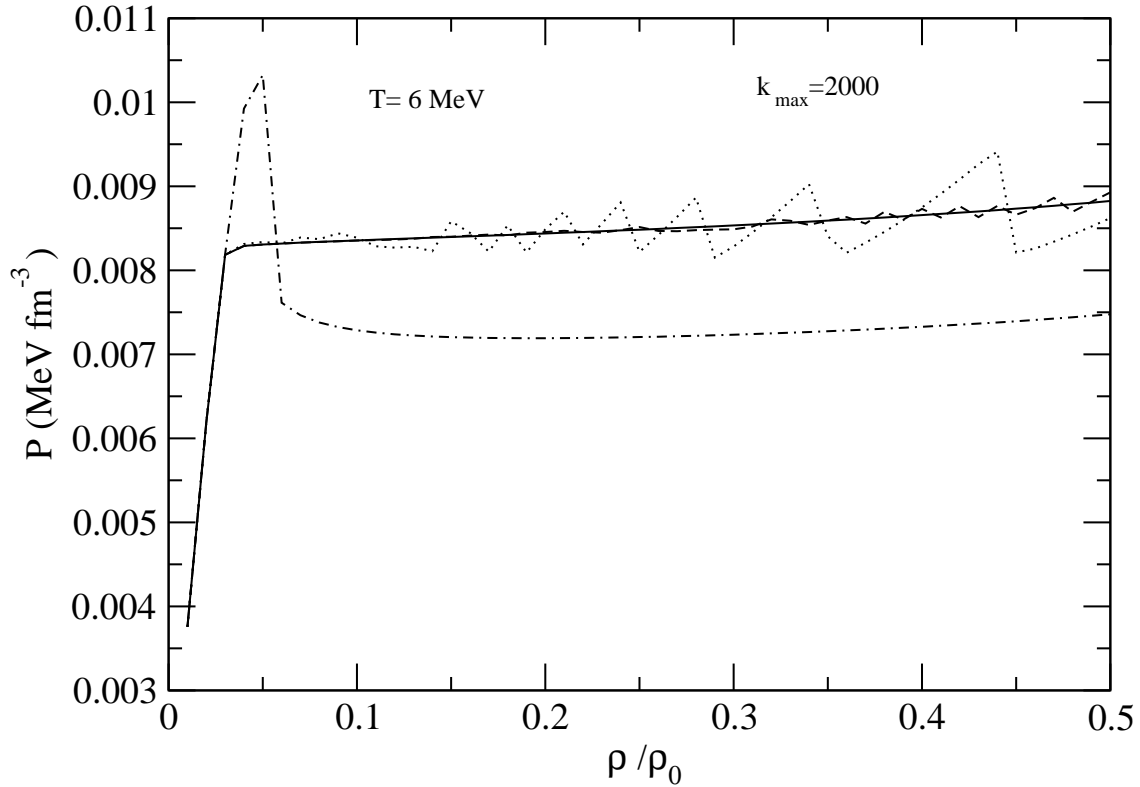}}
\end{center}
\caption{ Pressures calculated in the canonical model compared with
pressure calculated in the grand canonical model.  For all of these
the largest composite allowed has 2000 nucleons ($k_{max}=2000$)
and the temperature is 6 MeV.  The grand canonical calculation
is the solid curve.  The
canonical calculations are done with $N$=2000 (dash-dot), $N$=100,000
(dots) and $N$=500,000(dash).  As $N$ increases, agreement with the 
grand canonical result becomes better and better.}
\end{figure}
\pagebreak
\begin{figure}[htb]
\begin{center}
{\includegraphics[angle=0,width=12cm]{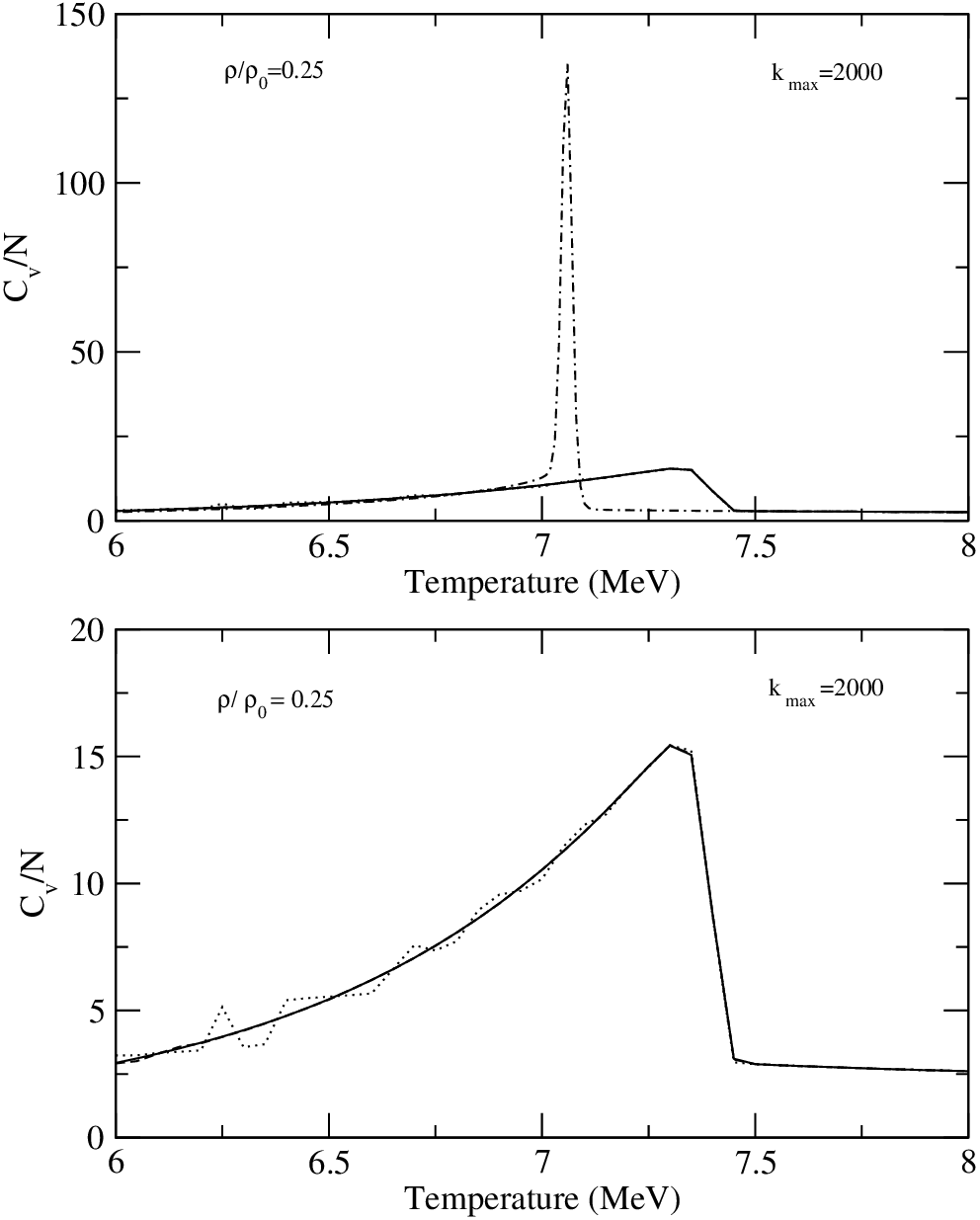}}
\end{center}
\caption{ Specific heat at constant volume in grand canonical and
canonical models.  Two different scales are needed to highlight
differences in values.  Again the canonical calculations are done 
with $N$=2000 (dash-dot), $N$=100,000 (dots) and $N$=500,000 (dash).
The difference between the grand canonical result (solid) and the $N$=2,000
calculation is huge around 7 MeV (upper panel) but for $N$=100,000
and $N$=500,000 the results are so close to the grand canonical
values that they are nearly indistinguishable in the scale of the
upper panel.  In the lower panel results for $N$=100,000 and $N$=500,000
are compared with grand canonical values.  Even in this expanded scale
the $N$=500,000 canonical results are indistinguishable in the curve
from the grand canonical results.} 
\end{figure}
\pagebreak
\begin{figure}[htb]
\begin{center}
{\includegraphics[angle=0,width=15cm]{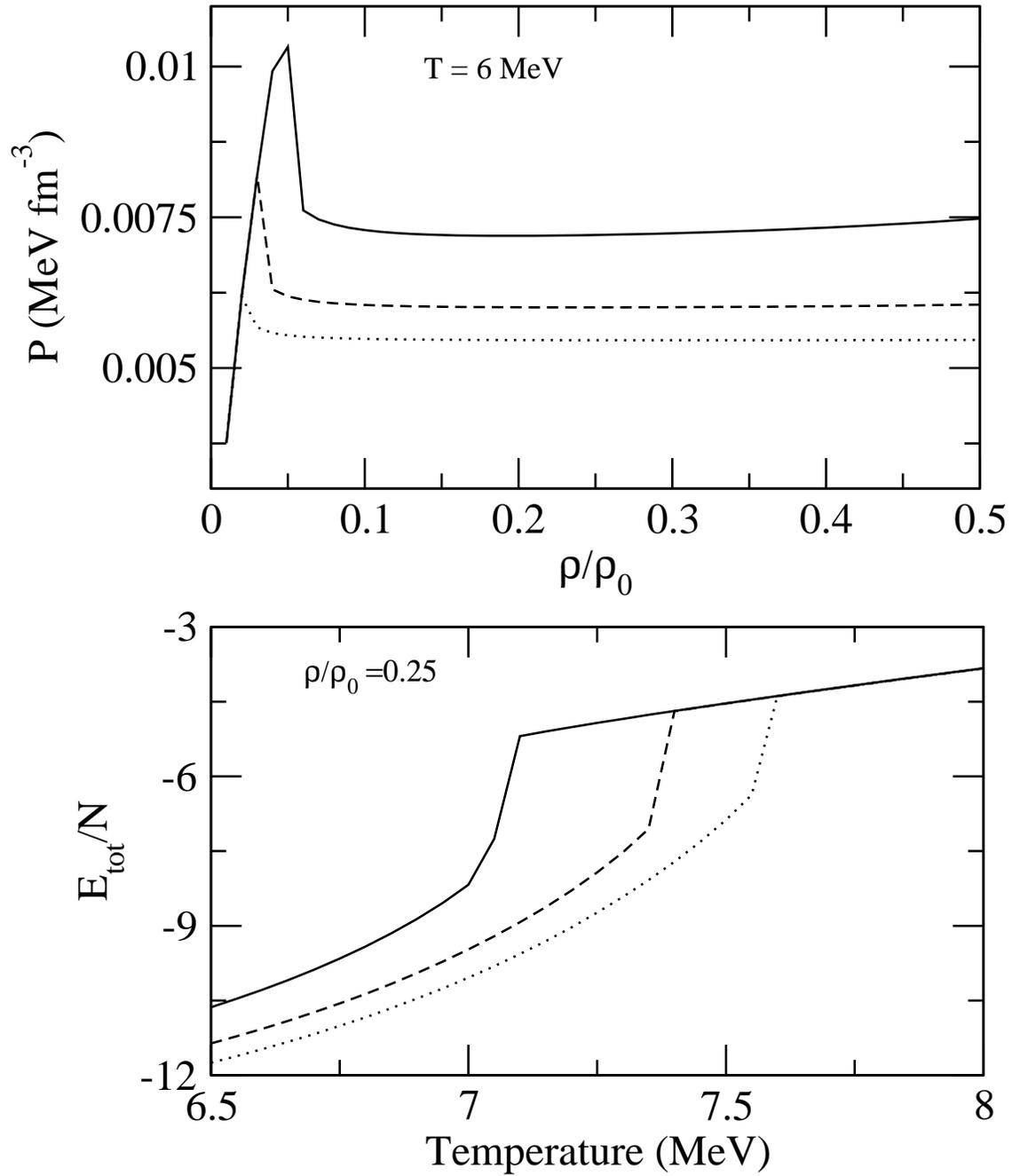}}
\end{center}
\caption{ Pressure and energy per particle for the canonical model
of $N=k_{max}$ for $N$=2000 (solid), N=10,000 (dash) and $N$=50,000
(dot).}
\end{figure}
\pagebreak
\begin{figure}[htb]
\begin{center}
{\includegraphics[angle=0,width=15cm]{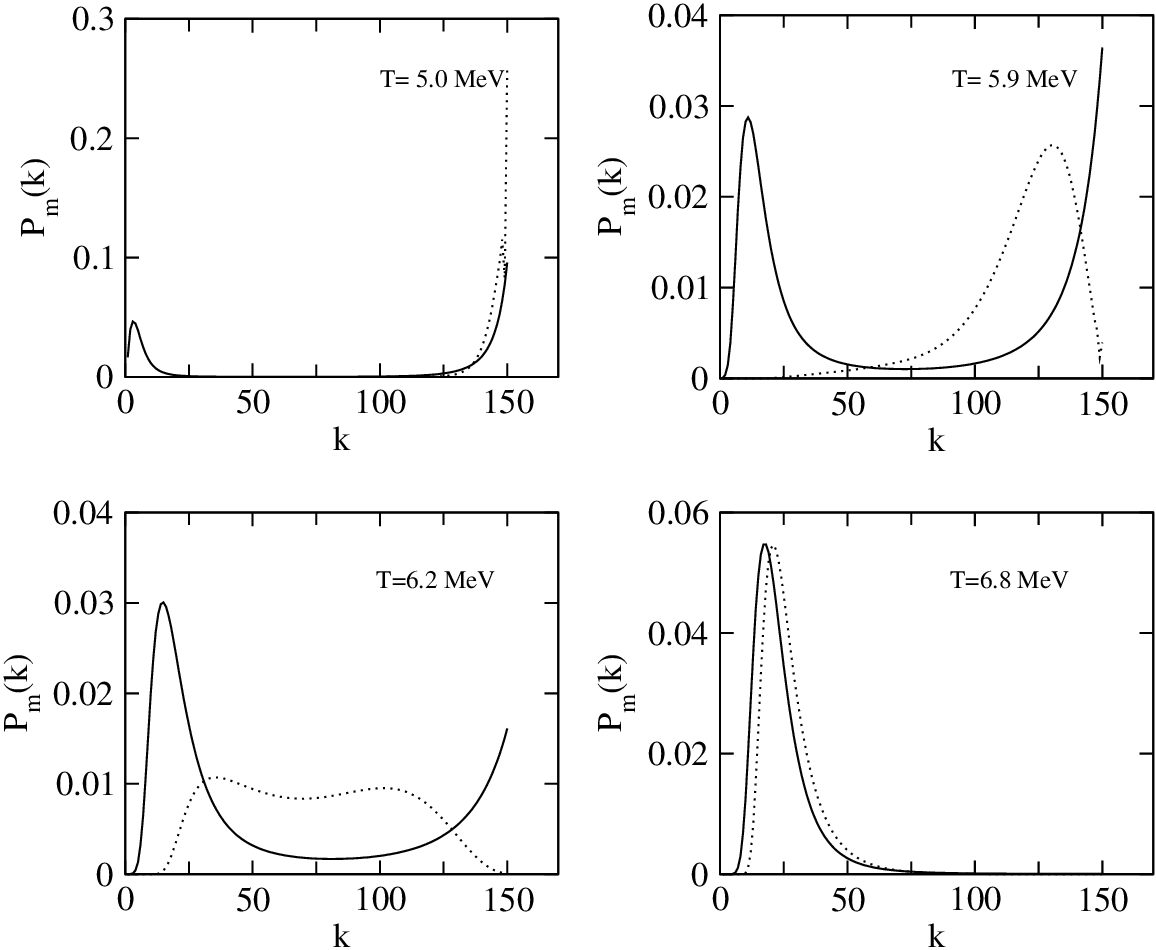}}
\end{center}
\caption{ Probability that the largest cluster has $k$ nucleons
plotted as a function of $k$ in the grand canonical(solid) and 
the canonical model(dot).
Here $N$=150 in the canonical model and in the grand canonical model
the average value is set at $N$=150.  The value of $k_{max}$ is also 150.
The density is fixed at $\rho/\rho_0=0.25$.
In the grand canonical model bimodality is seen at about 5.9 MeV and in 
the canonical model this appears at about 6.2 MeV.}
\end{figure}
\pagebreak
\begin{figure}[htb]
\begin{center}
{\includegraphics[angle=0,width=15cm]{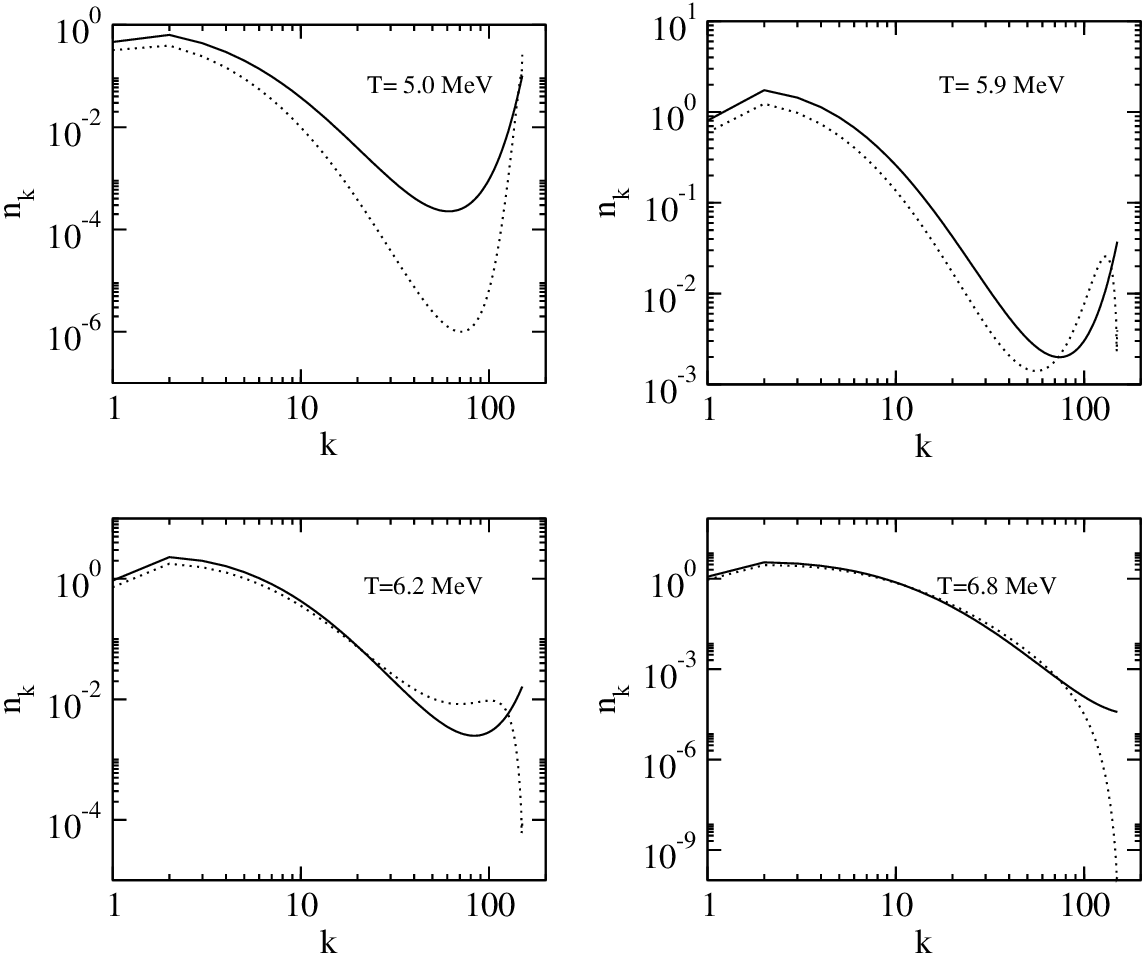}}
\end{center}
\caption{ For the same cases as above, 
the average multiplicity of each composite plotted as a
function of mass number $k$.  Note that the grand canonical results
(solid) approximate the canonical results (dots) quite well at the highest
temperature (except at very high mass numbers) but the agreement worsens
as the temperature is lowered.}

\end{figure}

\end{document}